\documentclass[preprint,eqsecnum,aps]{revtex4}
\usepackage{graphicx}
\begin{document}
\title{Scalar kinks and fermion localisation in warped spacetimes}
\author{Ratna Koley \footnote{Electronic address: {\em ratna@cts.iitkgp.ernet.in}}
${}^{}$and Sayan Kar \footnote{Electronic address : {\em sayan@cts.iitkgp.ernet.in}
}${}^{}$}
\affiliation{Department of Physics and Centre for
Theoretical Studies \\Indian Institute of Technology, Kharagpur 721 302, India}
\begin{abstract}
Scalar kinks propagating
along the bulk in warped
spacetimes provide a thick brane realisation of the braneworld. 
We consider here,
a class of such exact solutions of the full Einstein-scalar system with a
sine-Gordon
potential and a negative cosmological constant.
In the background of the kink 
and the corresponding warped geometry, we 
discuss the issue of localisation of
spin half fermions (with emphasis on massive ones) on the brane in the
presence of different types of kink-fermion Yukawa couplings.
We analyse the possibility of quasi-bound states for large values of 
the Yukawa coupling parameter $\gamma_F$ (with $\nu$, 
the warp factor parameter kept fixed) using appropriate, recently developed, 
approximation methods. In particular, the spectrum of the low--lying states and their lifetimes are obtained, with the latter being exponentially
enhanced for large $\nu \gamma_F$. 
Our results indicate quantitatively, within this model, that it is
possible to tune the nature of warping and the strength and form of the Yukawa
interaction to obtain trapped massive fermion states 
on the brane, which, however, do have a finite (but very small) probability  
of escaping into the bulk.    

\end{abstract}
\maketitle


\section{Introduction}

Extra dimensions {\em a la} Kaluza--Klein {\cite{kk,kkref} have been
around as an elegant theoretical construct for decades. Their
inevitable presence in superstring theories {\cite{string}}, is by now,
well-established.
Of late,
warped (nonfactorisable) spacetimes have been in vogue ever
since Randall--Sundrum
introduced them in the context of the now--popular braneworld models {\cite{rs}.
In such models, extra dimensions exist, can be compact or non--compact, but
the usual schemes of `compactification' are abandoned.
The four dimensional world (the so--called 3--brane)
is visualised as an embedded hypersurface in the five
dimensional bulk. Effective field theories on the 3--brane are
obtained without invoking `compactification' in the usual sense
of Kaluza--Klein. Further details of how this is achieved can be
found in {\cite{rs}} or in subsequent review articles {\cite{rsrev}}.
Much work has been carried out
on diverse aspects of such models in the last four years.
These include the attempted resolution of the heirarchy problem
{\cite{besan,rs}}, questions about the
localisation of various types of fields on the brane {\cite{bajc}},
particle phenomeonology in the braneworld context {\cite{besan}}
and cosmological consequences {\cite{cosmo}}.
The actual existence of warped extra dimensions as well as a firm
foundational basis for these models however, still remain open issues.
Phenomenological
work supported by experimental verification and simultaneous progress in
formal theory (via superstrings or any other yet-to-be-discovered
fundamental unified theory)
will perhaps be able to provide
conclusive answers about the future of the braneworld scenario.

In the early braneworld models, 
the
bulk spacetime was assumed
to have only a negative cosmological constant (anti-de Sitter space
in five dimensions). Subsequently, we have
had models with fields in the bulk--the simplest of them being those with
a bulk
scalar field dependent only on the fifth (extra) coordinate
{\cite{bulk}}. 
In this paper, we first consider exact solutions of the Einstein--scalar
equations  with a sine-Gordon potential. There exists a
full solution of this system with the scalar field configuration being a
kink. Such a configuration, the so--called thick brane,
provides a realisation of the 
braneworld as a domain wall in the bulk, a fact which has been 
illustrated in several examples in the literature {\cite{ringeval}}. 
The warped background spacetime, in which
the bulk sine-Gordon soliton is a self--consistent solution, turns out to
have a non--constant but asymptotically negative Ricci curvature. 

In the scenario mentioned above, where we have an exact solution
of the full Einstein-scalar system, we address the question of 
localisation of fermions with emphasis on the massive ones.
Though largely a toy model, we are able to find some interesting 
analytical results, which add to the only work on massive fermions
in the literature {\cite{ringeval}}.
To address the issue of localisation, we couple the
fermion field with the scalar field through a Yukawa coupling, as has been
done in many articles on fermion localisation.
Subsequently, we investigate and compare the localisation effects  
for two different types of Yukawa couplings 
in the background of this scalar sine--Gordon kink
in the warped line element. Known results for the massless fermions are
verified. Using approximate methods, we are able to delineate which
Yukawa coupling can yield better results (i.e. have more trapped states).
In addition, we make use of a new, recently developed formalism for estimating
the lifetime of the states and show how the lifetime can indeed be exponentiallylarge for large values of the Yukawa coupling parameter and moderate
values of the warp factor parameter.
In summary, the available exact solution for the scalar field and the
warp factor helps us to understand analytically, the nature of the
effective potential, the trapped states, their lifetime and the 
comparative role of different Yukawa couplings.   

\section{
The exact solution for the warp factor and the SG kink} 

We begin by writing down the Lagrangian and the equations of motion
for Einstein gravity with a cosmological constant,
minimally coupled to a real scalar field with
a potential $V(\phi)$. The action for such a system is given as :

\begin{equation}
S = \int \left [ \frac{1}{2\kappa_5^2}\left (R-2\Lambda\right ) -\frac{1}{2}
g^{ab}\partial_a \phi \partial_b \phi - V(\phi) \right ] \sqrt{-g} d^5 x
\end{equation}

where $g_{ab}$ is the five dimensional metric tensor with signature
(- + + + +) and R is the Ricci scalar. $\Lambda$ is the five
dimensional cosmological constant. $\kappa_5^2=8 \pi G_5$ where
$G_5$ is the five dimensional Newton constant.

We assume a line element, in the five dimensional bulk, of the form :

\begin{equation}
ds^2 = d\sigma^2 +e^{-2f(\sigma)} \eta_{ij}dx^i dx^j
\end{equation}

where $\sigma$ is the fifth coordinate and $f(\sigma)$ is the
warp factor. The scalar field is considered to be a function of
$\sigma$ only. With these ansatze the Einstein scalar system
reduces to the following system of coupled, nonlinear ordinary
differential equations :

\begin{eqnarray}
f '' & = & a {\phi'}^2 \\
{f'}^2 & = & \frac{a}{4} \left ( {\phi'}^2 - 2V \right ) -\frac{\Lambda}{6}\\
\phi'' -4f'\phi' &  = & \frac{dV}{d\phi}
\end{eqnarray}

where $a=\kappa_5^2/3 $.
The first two of the above set are the Einstein equations and the third is
the scalar field (Klein-Gordon) equation, which follows from the Einstein
equations and is not independent.

The obvious question is : is there an exact
analytical solution to the system of Eqns (2.3)--(2.5)? For the Higgs
potential, Ringeval et al {\cite{ringeval}}
have numerically
investigated the solutions and found kinks and the corresponding
warp factors. We shall now show that the Eqns (2.3)-(2.5) can indeed
be solved exactly for a sine-Gordon potential ($V(\phi)=B\left (1+
\cos \frac{2\phi}{A}\right )$. 
Variants of this solution with
different choices of $V(\phi)$ have appeared in the literature {\cite{gremm}}.
We comment on this at the end of this section. 
We now write down the solution explicitly below.

\begin{equation}
f (\sigma) = \frac{a}{\kappa_1} \sqrt{\frac{\vert\Lambda \vert}{6}}\ln
\cosh \left (\frac{\kappa_1}{a}\sigma\right )
\end{equation}

\begin{equation}
\phi(\sigma) = 2A \tan^{-1}\left ( \exp{\frac{\kappa_1}{a}\sigma} \right )
-\frac{\pi A}{2}
\end{equation}

The constant B in the sine-Gordon potential is given as :

$B= \frac{\vert \Lambda\vert}{6a^2} \left (a+\frac{1}{4A^2}\right )$.
Also $\kappa_1 = \frac{1}{A^2}\sqrt{\frac{\vert \Lambda \vert}{6}}$.

It is important to note the presence of a negative cosmological constant
in both the warp factor and the soliton. In addition, we emphasize that
the kink/soliton is constructed in such a way that $\phi(\sigma)$ is an
odd function about the origin. The second term in the expression for
$\phi(\sigma)$ is required to achieve this. 

This model with a bulk SG potential provides a `thick brane' realisation
of the Randall--Sundrum scenario where the SG field and its soliton
configuration dynamically generate this domain wall configuration
in the background warped geometry. In addition, as is obvious from the
functional from of $f$, there is no discontinuity in the derivative of
$f$ at the location of the brane. The warp factor is smooth everywhere
and has all the necessary features.    
 
The line element in the warped bulk space time is given by

\begin{equation}
ds^{2} = d\sigma^{2} + \cosh^{-2\nu} (\frac{\kappa_1}{a}\sigma)  \eta_{ij}dx^i dx^j
\end{equation}

where $\nu = \frac{a}{\kappa_1} \sqrt{\frac{\vert\Lambda \vert}{6}}$

Notice that the metric is completely non-singular for the full domain of
the fifth coordinate. It describes a space of negative Ricci curvature
given by

\begin{equation}
R= \frac{\vert \Lambda \vert}{6} \left [ \frac{8}{aA^2} -
\left (\frac{8}{aA^2}+20 \right ) \tanh^2 \frac{\kappa_1}{a} \sigma
\right ]
\end{equation}

It is straightforward to check that the above function for R is
not always negative with a asymptotic ($\sigma \rightarrow \pm \infty$)
value of $-\frac{10}{3}\vert \Lambda
\vert$. One cannot get {\em constant} negative Ricci curvature for any real
value of the parameter $A$. In the above solution,
the parameter $A$ is free whereas
$B$ is fixed in terms of $A$, $\vert \Lambda \vert$ and $a$.

Furthermore, since the above $f(\sigma)$ and $\phi(\sigma)$ consistently
solves the Einstein-scalar system we may want to know precisely the
energy density and pressures which generates such a line element.
These are given below.

\begin{equation}
\rho = -p_{x,y,z}=3\frac{\kappa_1}{a} \sqrt{\frac{\vert \Lambda \vert}{6}}
\mbox{sech}^2\frac{\kappa_1}{a}\sigma - \vert \Lambda \vert \tanh^2\frac{\kappa_1}{a}
\sigma
\end{equation}
\begin{equation}
p_{\sigma}= \vert \Lambda \vert \tanh^2\frac{\kappa_1}{a}\sigma
\end{equation}

Note that, asymptotically, the stress energy is that of anti-deSitter space
in five dimensions. At $\sigma=0$ (i.e. the location of the brane)
we have $p_{x,y,z}=-\rho$ and $p_{\sigma}=0$ -- an effective cosmological
constant on the 3--brane. The energy density and pressures are plotted in
Figure 2. It is worth noting that the matter stress-energy which acts
as a source for the warped geometry satisfies the Null Energy Condition
(NEC $\rightarrow$  $ \rho+p_i\ge 0$) though the Weak and Strong
 Energy Conditions (WEC and SEC) are violated 
(WEC $\rightarrow$ $\rho \ge 0, \rho+p_i \ge 0$) {\cite{visser}}. 

\begin{figure}
\includegraphics[width= 10cm,height=5.5cm]{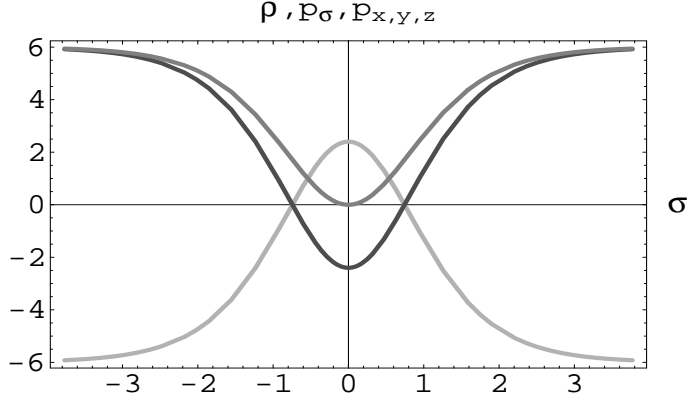}
\caption{Variation of the energy density  $\rho$ (gray line) and pressure ( $p_i$ (black line) and 
 $p_\sigma$ (dark gray line)) with $\sigma$ corresponding to the warped geometry given by the 
metric $ (4.3) $}. 
\end{figure}

It is worth noting that there exists related work on similar solutions
for $\phi$ and $f$. In a most general case one might consider 
the following :

\begin{equation}
f(\sigma) = A_1 \ln (\cosh b\sigma) \hspace{.1in}; \hspace{.1in}
\phi (\sigma) = 2 A_2 \mbox{tan}^{-1} (\mbox{exp}(b\sigma)) -\frac{\pi A_2}{2}\hspace{.1in}
;\hspace{.1in} V(\phi) = C + D \cos (2\frac{\phi}{A_2})
\end{equation}

Using these in the equations of motion one obtains the following
constraints on $A_1$, $A_2$, $C$, $D$ and $\vert \Lambda \vert$.

\begin{eqnarray}
A_1 = a A_2^2 \hspace{.1in}; \hspace{.1in} A_2^2 b^2\left (1+4 A_1^2\right )
=8B \\
C= -\frac{A_1^2 b^2}{a} + \frac{A_2^2 b^2}{4} +\frac{\vert \Lambda \vert}{3}
\\
D= \frac{A_1^2 b^2}{a} + \frac{A_2^2 b^2}{4} 
\end{eqnarray}

It is easy to note from the above that if $\Lambda=0$ it is not possible to
have $C=D$. Various expressions for $V(\phi)$ with the corresponding solutions
have been noted in the literature {\cite{gremm}} though these are all for
$\Lambda =0$. The solutions for $\vert \Lambda \vert =0$ and $A_1\neq A_2$
can be matched with the solutions for $\vert \Lambda \vert \neq 0$ with
$A_1=A_2=2D$, if, in the former, we assume $A_2-A_1 = 2\frac{\vert \Lambda 
\vert}{3a}$. This means that an `effective' $\Lambda$ will appear within
the scalar potential. However, we prefer to keep $\Lambda$ separate. As we
shall see later, the mass spectrum of the localised fermions will be
dependent on $\eta_F$ and $\vert\Lambda \vert$. It is also worth
mentioning that in some of the literature on similar domain wall
solutions the SG soliton is taken to be a neither-odd nor-even function.
We have used a shifted version (shift along the $\phi$ axis, in order
to maintain the odd character of the soliton solution and also to
have $\phi=0$ at the brane location). The use of the purely odd
soliton is important because, as we note in the next section, the
symmetry of the effective potential for massive fermions w.r.t. the
brane location is respected only if $\phi$ is purely odd. This is
desirable because the physics on either side of the brane should have
identical features ($Z_2$ symmetry).

In this context, one  must also mention the extensive work done in the recent past
on non-supersymmetric as well as supersymmetric {\cite{nonsusy}},
{\cite{susy}} domain walls in diverse models and theories.

Let us now focus on the question of
localisation of massive and massless fermion fields on the 3--brane.

\section{Fermion localisation}

As with other fields such as scalars, spin one fields, gravitons, gravitinos,
it is a pertinent question to ask whether spin-half fermions are/can be localised 
on the brane. To address this issue, one needs to assume fermions as propagating
in the bulk (i.e. spinor fields functionally dependent on
$\sigma$) in the five dimensional warped background geometry.
It has been shown {\cite{bajc,seif}} that massless fermions in a warped
geometry with a RS warp factor (proportional to $\sigma$) cannot be localised.
However with additional couplings with, say, a scalar (e.g. Yukawa
couplings) it is possible to obtain localised massless chiral fermions on the
brane following the method developed in {\cite{rebbi}} (in 4D flat spacetime).
The question of localisation of massive fermions is a problem which
we shall address here for the toy model with a sine-Gordon potential and
a Yukawa coupling between the fermion and the scalar fields.

The simplest set up for
localisation is as follows. As mentioned earlier, one can achieve
localised fermions on the brane by
introducing the Yukawa coupling between fermion field and the scalar field, $\eta_{F}
\bar{\Psi} \mbox{F}(\Phi) \Psi$ where $\mbox{F}(\Phi)$  is a
function of the rescaled scalar field (i.e. $\Phi =\frac{\phi}{A}$).
Let us first obtain the equation that governs the bulk motion of the fermions. To start
with, we find that the Lagrangian for a Dirac fermion propagating in a
five dimensional space with the metric (2.8), is given by

\begin{equation}
\sqrt{-g} {\cal{L}}_{Dirac} = \sqrt{-g}\hspace{.03in}(i \bar{ \Psi} \Gamma^{a} {\cal{D}}_{a} \Psi
- \eta_{F} \bar{ \Psi} \mbox{F}(\Phi) \Psi )
\end{equation}

where $g = \mbox{det}(g_{ab})$, is the determinant of full five dimensional metric. 
The matrices $\Gamma^{a} = (e^{f(\sigma)} \gamma^{\mu}, -i \gamma^{5})$ provide a 
four dimensional
representation of the Dirac matrices in five dimensional curved space. Where $\gamma^{\mu}$
and $\gamma^{5}$ are the usual four dimensional Dirac matrices in chiral representation.
The Clifford algebra in curved space, $\{ \Gamma^{a},\Gamma^{b} \}=2 g^{ab}$ is obeyed by the
five dimensional gamma matrices as their counterparts follow in four dimensions.
These representations have the useful property that
they can induce a chiral particle theory on the brane {\cite{rebbi}}.

The covariant derivative in 5D curved space can also be calculated for the metric given
in Eqn. (2.8) {\cite {wb}}:

\begin{equation}
{\cal{D}}_{\mu} =(\partial_{\mu} -\frac{1}{2} f'(\sigma) e^{-f(\sigma)} \Gamma_{\mu}
\Gamma^{4}) ; \hspace{.8cm}
{\cal{D}}_{4} = \partial_{\sigma}
\end{equation}

Applying the above configuration to the Eqn.(3.1) one obtains the Dirac Lagrangian in 5D
curved spacetime in the following form
 
\begin{equation}
\sqrt{-g} {\cal{L}}_{Dirac} = e^{- 4 f (\sigma)}  \bar{\Psi}\left[i e^{f(\sigma)}\gamma^{\mu}\partial_{\mu} + 
\gamma^{5} (\partial_{4} -2 f'(\sigma)) - \eta_{F} \mbox{F}(\Phi) \right ] 
\Psi  
\end{equation}

The dimensional reduction from 5D to 4D is performed in such a way that the
standard four dimensional chiral particle theory is recovered. The five dimensional 
spinor can be decomposed into four dimensional and 
fifth dimensional parts: $\Psi(x^{\mu},\sigma) =
\Psi(x^{\mu}) \xi(\sigma)$. Since the four dimensional massive
fermions require both the left and right chiralities
it is convenient to organise the spinors with respect to
$\Psi_{L}$ and $\Psi_{R}$ which represent four component spinors living in five
dimensions given by $\Psi_{L,R} = \frac{1}{2} (1 \mp \gamma_{5})\Psi$.
Hence the full 5D spinor can be split in the following way

\begin{equation}
\Psi(x^{\mu},\sigma) = \left( \Psi_{L}(x^{\mu})\xi_{L}(\sigma) +
\Psi_{R}(x^{\mu})\xi_{R}(\sigma) \right)
\end{equation}

where we have to determine $\xi(\sigma)$ which is an eigenfunction 
dependent only on $\sigma$. Let us assume that the functions $\xi_{L.R}(\sigma)$ 
satisfy the following eigenvalue equations

\begin{eqnarray}
 e^{-f(\sigma)}\left [\partial_{\sigma}-2 f'(\sigma) - \eta_{F} \mbox{F}(\Phi) \right ] \xi_{R}(\sigma)
& = & -m \xi_{L}(\sigma) \\
 e^{-f(\sigma)} \left [\partial_{\sigma}-2 f'(\sigma) + \eta_{F} \mbox{F}(\Phi) \right ] \xi_{L}(\sigma)
& = & m \xi_{R}(\sigma)
\end{eqnarray}

The full 5D action then reduces to the standard four dimensional action 
for the massive chiral fermions (with mass $m$) , 
when integrated over the extra dimension \cite{ringeval}, provided 
(a) the above equations are satisfied by the bulk 
fermions and (b) the following orthonormality conditions are obeyed.

\begin{eqnarray}
\int_{-\infty}^{\infty} e^{-3 f(\sigma)} \xi_{L_{m}} \xi_{L_{n}} d\sigma =
\int_{-\infty}^{\infty} e^{-3 f(\sigma)} \xi_{R_{m}} \xi_{R_{n}} d\sigma = \delta_{m n} \\
\int_{-\infty}^{\infty} e^{-3 f(\sigma)} \xi_{L_{m}} \xi_{R_{n}} d\sigma = 0
\end{eqnarray}

The Yukawa coupling between the scalar and the fermion, 
with the kink solution for the scalar,
is necessarily like an effective, variable, 5D mass 
for the fermions \cite{bajc}. This is largely responsible for
generating the massive fermion modes in four 
dimensions. The dynamical features of the model 
can thus be obtained from the solutions of the eigenvalue equations 
(3.5) and (3.6).     

Let us first focus on massless (i.e. $m = 0$) fermions for a Yukawa coupling of the form
$\eta_{F} \bar{\Psi} \Phi \Psi$ (i.e. $\mbox{F}(\Phi)= \Phi$). In this case Eqn. 
(3.5) and (3.6)
reduce into two decoupled equations. The solutions of which are asymptotically in the
following form

\begin{eqnarray}
\xi_{L}(\sigma) & = & e^{-\left (\frac{\eta_{F}\pi}{2}  - 2 \sqrt{\frac{\vert \Lambda \vert}{6}}
\right ) \vert \sigma \vert} \\
\xi_{R}(\sigma) & = & e^{\left (\frac{\eta_{F} \pi}{2}  + 2 \sqrt{\frac{\vert \Lambda \vert}{6}}
\right ) \vert \sigma \vert}
\end{eqnarray}

Eqn. (3.9) yields  the localisation of left chiral fermions on the brane so long as,
$\eta_{F} \ >  \frac{4}{\pi} \sqrt{\frac {\vert \Lambda \vert}{6}}$.
We can as well achieve the bounded right chiral states by
considering the interaction with an anti-kink sine-Gordon profile.

We now turn our attention towards the massive fermions i.e. $m \not= 0$.
Defining  $\tilde\xi_{L.R}(\sigma) = e^{-\frac{3}{2} f(\sigma)} \xi_{L.R} (\sigma)$
and making use of $\partial_{\sigma}\tilde\xi_{R}(\sigma)$,
$\tilde\xi_{R}(\sigma)$ from Eqn. (3.5) and (3.6) we obtian
the following second order, decoupled equation for the 
left chiral fermions whereas the right chiral states can be
completely defined from the later equation with a prior knowledge to
$\tilde\xi_{L}(\sigma)$ :

\begin{eqnarray}
& & \left [\partial^{2}_{\sigma} - 2f'(\sigma) \partial_{\sigma} +
m^{2} e^{2f(\sigma)} + \frac{3}{4} f'^{2}
(\sigma) - \eta_{F}^{2} \mbox{F}(\Phi)^{2} \right .  \nonumber \\
& - & \left . \eta_{F} f'(\sigma)
\mbox{F}(\Phi) + \eta_{F} \frac{\mbox{dF}}{\mbox{d}\Phi} \Phi'(\sigma)
-\frac{f''(\sigma)}{2} \right ] \tilde\xi_{L}(\sigma)  =  0\\
& & \tilde\xi_{R}(\sigma)  = \frac{e^{-f(\sigma)}}{m} \left [\partial_{\sigma} + \eta_{F}
\mbox{F}(\Phi) - \frac{f'(\sigma)}{2}\right ]\tilde\xi_{L}(\sigma)
\end{eqnarray}

We note that the equation satisfied by $\tilde\xi_{L} (\sigma)$ (Eqn. 3.11) is
a second order
differential equation that looks like a time--independent 
Schr\"{o}dinger equation (more precisely like a `zero energy'
Schr\"{o}dinger equation), when recast in the following form

\begin{equation}
\partial^{2}_{\sigma} \hat\xi_{L}(\sigma) + \left[m^{2}e^{2f(\sigma)} +
\frac{f''(\sigma)}{2} + \eta_{F} \frac{\mbox{dF}}{\mbox{d}\Phi}
\Phi'(\sigma)-(\eta_{F}\mbox{F}(\Phi) + \frac{f'(\sigma)}
{2})^{2}\right ]
\hat\xi_{L}(\sigma) = 0
\end{equation}

where,  $\hat\xi_{L}(\sigma) = e^{-f(\sigma)}\tilde\xi_{L}(\sigma)$.

We are interested in localisation of massive fermions around the brane. 
The effective potential for the left chiral fermions  
obtained from the Eqn. (3.13) should have a minimum at the brane in order to 
ensure the confinement of
fermions.  The $Z_{2}$ symmetry of the bulk spacetime with respect to
the position of the brane (at $\sigma=0$) puts a restriction on the
choice of the form of the Yukawa coupling.
If we demand such a symmetry for the effective potential 
$F(\Phi)$ should necessarily be an odd function of $\Phi$.
 For example,
if we consider the coupling as  $\eta_{F} \bar{\Psi} \Phi \Psi$   (i.e. $\mbox{F}(\Phi)$ 
= $\Phi$) one obtains a symmetric effective potential given by

\begin{eqnarray}
V_{eff}(y) & = & \left (2 \gamma_{F} \tan^{-1} (e^{\hat{y}})- \frac{\pi\gamma_F}{2} +\frac{1}{2}
\sqrt{\frac{1}{6}} \tanh(\hat{y})\right )^{2}
 - \frac {\kappa_{1}}{2 a} \sqrt{\frac{1}{6 \vert \Lambda \vert}} \mbox{sech}^{2} (\hat{y})
\nonumber \\
& - & \frac{\gamma_{F}}{\sqrt{\vert \Lambda \vert}} \frac{\kappa_{1}}{a} \mbox{sech} (\hat{y})
-  \mu^2 \cosh^{2 \nu} (\hat{y})
\end{eqnarray}

The above relation is written using a dimensionless variable and
dimensionless parameters :

\begin{equation}
y = \sigma \sqrt{\vert \Lambda \vert} \hspace{.2in};\hspace{.2in}
\mu = \frac{m}{\sqrt{\vert \Lambda \vert}} \hspace{.2in} ; \hspace{.2in}
\gamma_F = \frac{\eta_F}{\sqrt{\vert \Lambda \vert}}
\end{equation}

and for simplicity $\frac{\kappa_{1}}{ {a \sqrt{\vert
\Lambda \vert}}} y$ has been renamed as $\hat{y}$. 

\begin{figure}[htb]
\includegraphics[width= 10cm,height=5.5cm]{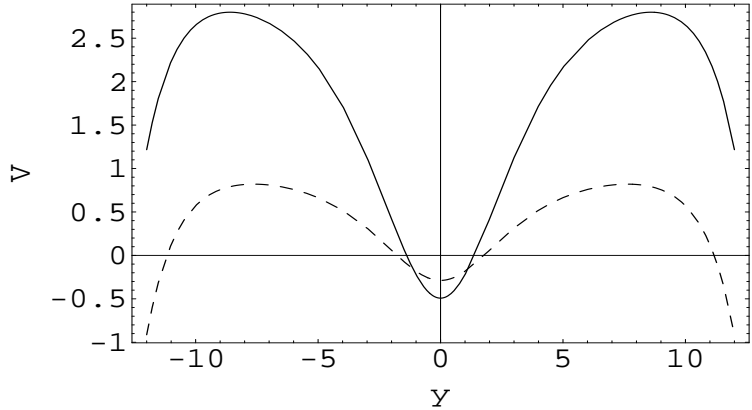}
\caption{For $\mbox{F}(\Phi) = \Phi$, the effective potential V is plotted as a function of the
rescaled extra dimension, y for different values of $\gamma_F$. The black curve
corresponds to $\gamma_{F} = 1$, whereas the dashed curve represents V for
 $\gamma_{F} =\frac{1}{2}$.
The parameters are chosen as
$\nu = 1$
and $ \Lambda = -6 $.}
\end{figure}

Similarly, for $\mbox{F}(\Phi) = \sin\Phi$ the  $Z_{2}$
symmetric effective potential turns out to be :

\begin{eqnarray}
V_{eff}(y) & = & \left (\gamma_{F}  +\frac{1}{2}
\sqrt{\frac{1}{6}} \right )^{2}\tanh^{2}(\hat{y})
 - \frac {\kappa_{1}}{a \sqrt{\vert \Lambda \vert}} \left (\gamma_{F}  +\frac{1}{2}
 \sqrt{\frac{1}{6}} \right )
  \mbox{sech}^{2} (\hat{y})
-  \mu^2 \cosh^{2 \nu} (\hat{y})
\end{eqnarray}

\begin{figure}[htb]
\includegraphics[width= 10cm,height=5.5cm]{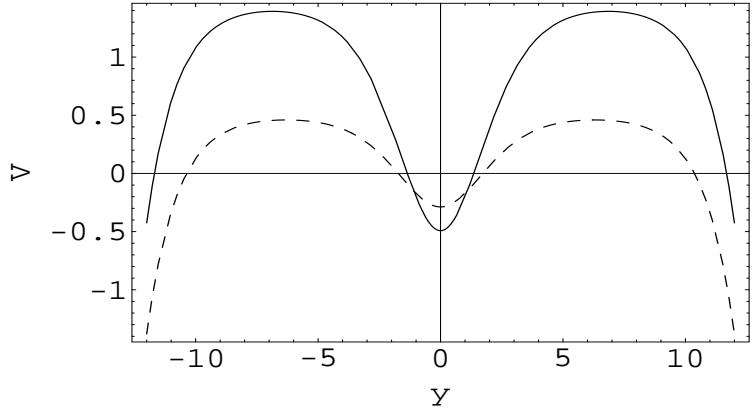}
\caption{The effective potential V is plotted as a function of the
rescaled extra dimension, y with different values of $\gamma_F$ for $\mbox{F}(\Phi) = \sin\Phi$.
 The black curve
corresponds to $\gamma_{F} = 1$, whereas the dashed curve represents V for
 $\gamma_{F} = 1/2$.
The parameters are chosen as
$\nu =1 $
and $ \Lambda = -6 $.}
\end{figure}

As mentioned before, the crucial point here is the antisymmetric
nature of the function $\mbox{F}(\Phi)$ under the transformation
 $\Phi \rightarrow -\Phi$. It can be be easily
shown that a symmetric $\mbox{F}(\Phi)$, such as $\cos\Phi$, $\Phi^2$
etc. 
yields an asymmetric effective potential. 
For the two different types of Yukawa interactions the 
effective potential acting on the left chiral fermions is shown in
Fig.(3) and (4) for varying values of $\gamma_F$. We have
chosen a value of $\mu$ which may not correspond to the
bound state. All the plots
show a minimum at $y =0$, which indicates that the
massive fermions can be trapped on the brane.
We notice, in addition, that the potential is, in general
symmetric about the point $y=0$. Furthermore, it is worth noting that
the value of the potential at $y = 0$ in both the cases is given by :

\begin{equation}
V_{eff}(y=0) = - \frac{\kappa_1}{a} \frac{1}{\sqrt{\vert \Lambda \vert}}
(\gamma_F + \frac{1}{2 \sqrt{6}}) -\mu^2
\end{equation}

However,
in either case, with increasing $\gamma_F$, the
depth of the well at the location of the brane becomes larger,
demonstrating that more trapped states are possible for stronger Yukawa
coupling. Additionally, 
the depth also increases
for small $\nu$ (or large $p$).  
On the other hand, for small $\gamma_F$ the depth is less. 
and we expect very few trapped states (maybe none). The nature of the
potential has prompted researchers to name it as a `volcano' potential. 
A fair amount of work related to bound states and resonances in such
volcano potentials has appeared in the literature in the recent past {\cite{volcano}}. 

We can obtain an exact solution of the equation (3.13) for $\gamma_F = 0$
(${\hat \xi}_L \sim \frac{\exp{(i\sqrt{A_2}r)}}{(1+r^2)^{\frac{1}{4}}}$ where
$r$ and $A_2$ are defined later).
However, this turns out to be an asymptotically growing 
function (recall that ${\tilde \xi}= e^{f(\sigma)}{\hat\xi}$ and we need to
consider $\tilde \xi$ or $\xi$ and not $\hat \xi$).
The function $\tilde\xi_{L}$ is also 
not normalisable according to the orthonormality conditions given in Eqn. (3.7). 
Based on all the above arguments we may therefore conclude 
that the confinement of massive fermions on the brane 
cannot be achieved without a strong Yukawa interaction.
Moreover the bound on the coupling constant, $ \eta_{F} > \frac{4}{\pi} \sqrt{{\vert \Lambda \vert}/6}$, 
required for the confinement of massless fermions also suggests 
that a large $\gamma_F$ is crucial for the existence of massive bound states on the brane.

We may also view the appearance and total number of bound states
as a result of a competition between
gravity and the Yukawa interaction. Without gravity, in a
flat five dimensional spacetime the effective potential reduces to the
P\"{o}sch-Teller potential,
 for which exact bound state solutions \cite{Landau} 
exist. But without the Yukawa interaction, we tend to get
very few bound states and mostly quasilocalised ones 
which may not be normalisable. Thus, the
presence of gravity, in some sense, destroys the bound states which were
there and therefore, a strong Yukawa
interaction is necessary in order to get them back. 
We shall now discuss these issues in some more detail.

\subsection{The mass spectrum of the low-lying states for $\mbox{F}(\Phi) = \Phi$ and  $\sin \Phi$ }

We expand the effective potential about
$y = 0$ and retain terms upto order $y^2$. This is the harmonic
oscillator approximation in the neighborhood of the 3--brane. The
second order differential equation for left chiral fermions in both the cases turns out
to be :

\begin{equation}
\partial^{2}_{y} \hat\xi_{L}(y) + \left[ \mu^2 -C_1
-\frac{1}{2} k y^2
\right ]
\hat\xi_{L}(y) = 0
\end{equation}

where we have for  $\mbox{F}(\Phi) = \Phi$:

\begin{eqnarray}
C_1 = -\left (\gamma_{F} + \frac{1}{2 \sqrt{6}} \right ) \frac{\kappa_{1}}{a \sqrt{\vert 
\Lambda \vert}} \\
C_2 =  \left (\gamma_{F} + \frac{1}{2 \sqrt{6}} \right )^2
\left (\frac{\kappa_{1}}{a \sqrt{\vert \Lambda \vert}} \right)^2 +
\left (\frac{\gamma_{F}}{2} + \frac{1}{2 \sqrt{6}} \right )
\left (\frac{\kappa_{1}}{a \sqrt{\vert \Lambda \vert}} \right )^3\\
k = 2 \left (C_2-\mu^2\frac{\kappa_1}{a}\frac{1}{\sqrt{6 \vert \Lambda \vert}}
\right )
\end{eqnarray}

and for $\mbox{F}(\Phi) = \sin\Phi$,  $C_1$ is the same as before and $C_2$
is given by : 

\begin{equation}
C_2 =  \left (\gamma_{F} + \frac{1}{2 \sqrt{6}} \right )^2
\left (\frac{\kappa_{1}}{a \sqrt{\vert \Lambda \vert}} \right)^2 +
\left (\gamma_{F} + \frac{1}{2 \sqrt{6}} \right )
\left (\frac{\kappa_{1}}{a \sqrt{\vert \Lambda \vert}} \right )^3
\end{equation}

From the harmonic oscillator approximation we can estimate the allowed energy levels
 and thereby find out the
allowed fermion mass spectrum. The full five dimensional
function $\Psi(x^{\mu}, \sigma)$ will have a definite parity 
because of the overall symmetry of the problem under $\sigma
\rightarrow -\sigma$. Thus the right and left chiral wavefunctions will be
either odd or even. We note here that only the even states will
survive in the spectrum because of the required matching of the
wave-function and its derivatives at the brane location (see a discussion
on this in {\cite{ringeval}}). The possible values
of $\mu^2$ are hidden in the expression :

\begin{equation}
\mu_n^2 = C_1 +\left ( 2n+ 1\right ) \sqrt{C_2 -\mu_n^2
\frac{\kappa_1}{a} \frac{1}{\sqrt{6 \vert \Lambda \vert}}}
\end{equation}

where $n=0,2,4....$. 
 
The discrete mass spectrum for the trapped
fermions can be obtained clearly from
the following relations
\vspace{0.3 cm}

For, $F(\Phi) = \Phi$ :

\begin{equation}
\mu_n^2 = 
-{\bar \gamma_{F}}\frac{1}{\nu\sqrt{6}} - \frac{(2n+1)^2}{12 \nu} +
(n+\frac{1}{2})\sqrt{\left (\frac{2n+1}{6\nu}\right )^2 + \frac{2\bar\gamma_F}{
3\sqrt{6} \nu^2} +\frac{2{\bar\gamma_F}^2}{3\nu^2} + \frac{{\bar\gamma_F}}{
3\sqrt{6} \nu^3} + \frac{1}{36 \nu^3}}
\end{equation}

For, $F(\Phi) = \sin\Phi$ :

\begin{equation}
\mu_n^2 = 
-{\bar \gamma_{F}}\frac{1}{\nu\sqrt{6}} - \frac{(2n+1)^2}{12 \nu} +
(n+\frac{1}{2})\sqrt{\left (\frac{2n+1}{6\nu}\right )^2 +\frac{2\bar\gamma_F}{
3\sqrt{6} \nu^2} +\frac{2{\bar\gamma_F}^2}{3\nu^2} + \frac{2{\bar\gamma_F}}{
3\sqrt{6} \nu^3}}
\end{equation}

where $\bar\gamma_F=\gamma_F +\frac{1}{2\sqrt{6}}$.

\begin{figure}
\includegraphics[width= 10cm,height=5.5cm]{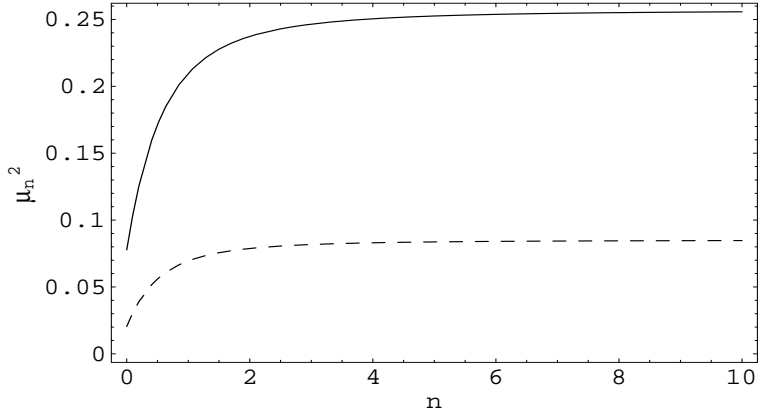}
\caption{The plot shows the variation of $\mu_{n}^2$ vs. n as given in
  Eqn. (5.34). Mass spectrum saturates very quickly 
for $\gamma_F = 0$, which has been shown for $\nu = 1/2$ (dashed curve) and 1 (continuous curve). 
The discrete levels reach to the continuum after a very few modes.}
\end{figure}

Notice the presence of the two parameters $\gamma_F$ and $\nu$. 
For $\gamma_F=0$ we find that the  
higher states quickly approach
the continuum. This is evident from the plot of $\mu_n^2$ versus
$n$ in Fig. (5) which shows that the difference between $\mu_n^2$ and
$\mu_{n+1}^2$ tends to saturate after the first few states. This behaviour
persists for nonzero but small $\gamma_F$ for which $\mu_n^2$ goes as
$\sqrt{\gamma_F}$. However, for large $\gamma_F$ the $\mu_n^2$ goes as
$\gamma_F$ and more bound states become available within the well.  

\subsubsection{Analysis for $\gamma_{F} >> 1$ and $F(\Phi) = \Phi$}

We now look for bounds on the trapped masses and the number of 
localised states in the limit $\gamma_F>>1$. The mass spectrum for
this case takes the form : 

\begin{equation}
\mu_n^2 = \frac{2 n  \gamma_F}{\nu \sqrt{6}}
\end{equation}

There is an upper limit on the mass of the fermions 
which are trapped by the effective
potential $(3.14)$ because it is necessary to have a potential barrier 
in order to have bound states. Let us check the asymptotic behaviour of $V_{eff}$.
For large values
of the extra dimension the effective potential reduces to :

\begin{equation}
V_{eff} \sim - \left (\mu^2 e^{\frac{2 y}{\sqrt{6}}} -
(\frac{\gamma_F \pi}{2} + \frac{1}{2 \sqrt {6}} )^2 \right )
\end{equation}

To have a potential which is a continuous function of $y$ one can match the approximate
effective potentials given in eqn. $(3.28)$ and $(3.37)$ and obtain the point of continuity
to be at $y = y_{m} \sim \frac{\pi}{2} \frac{a \sqrt{\vert \Lambda \vert}}{\kappa_1}$.
One finds the maximum
value of trapped mass from the expression $(3.37)$ 
using the following criterion for the vanishing of the barrier :

\begin{equation}
V_{eff} (\mu_{max},y_{m}) \sim 0
\end{equation}

Thus for $\gamma_{F} >> 1$, the maximum possible mass of the bound state is :

\begin{equation}
\mu_{max} = \frac{\gamma_F \pi}{2} \mbox{Exp} \left[- \frac{\pi a}{2 \kappa_1} \sqrt{\frac
{\vert \Lambda \vert}{6}}\right ] = \frac{\gamma_F \pi}{2}\exp{-(\frac{\pi}{2}
\nu )} 
\end{equation}

For example, choosing the parameters as given in Fig. $(3)$ the value
of $\mu_{max}$ $\sim$ $0.33 \gamma_F$. The maximum number of discrete massive states
localised around the brane is obtained easily from eqn. $(3.36)$ by introducing an extra $\frac{1}{2}$
factor to account for the even states only. For a particular choice 
of the
coupling constant and suitable range of parameters the value of $\mbox{n}_{max}$ can be 
calculated
from the integer obtained by the following relation :
 
\begin{equation}
\mbox{n}_{max} = \mbox{Int}\left [ \frac{1}{8} \left(\frac{ \gamma_F \pi^2}{2} \frac{a
\sqrt{\vert \Lambda \vert}}{\kappa_{1}}
\mbox{Exp} [-\frac{\pi a}{\kappa_1} \sqrt{\frac{\vert \Lambda \vert}{6}}  ] \right ) \right] 
= \mbox{Int} \left [ \frac{\sqrt{6}\gamma_F \pi^2}{16} \nu \exp {-(\pi \nu)}
\right ]
\end{equation} 

\subsubsection{Analysis for $\gamma_{F} >> 1$ and $F(\Phi) = \sin \Phi$} 

Let us now look at the case with $F(\Phi) = \sin \Phi$. The mass spectrum for 
very strong Yukawa coupling reduces to the form given in Eqn. (3.36).   
Far away from the brane the Effective potential turns out to be 

\begin{equation}
V_{eff} \sim - \left (\mu^2 e^{\frac{2 y}{\sqrt{6}}} -
(\gamma_F + \frac{1}{2 \sqrt {6}} )^2 \right )
\end{equation}

Following similar steps, one can easily obtain the potential barrier 
just disappears at the location, 
$y = \frac{a \sqrt{\vert \Lambda \vert}}{\kappa_{1}}$ along the extra dimension. 
Applying the matching condition given in Eqn. (3.38) one can readily obtain 
$\mu_{max}$ and $\mbox{n}_{max}$, which turn out to be :

\begin{eqnarray}
\mu_{max} = \gamma_F \mbox{Exp} \left[- \frac{a}{\kappa_1}\sqrt{\frac
{\vert \Lambda \vert}{6}} \right ] = \gamma_F \exp {(-\nu)}\\
\mbox{n}_{max} = \mbox{Int}\left [ \left(\gamma_F \frac{a
\sqrt{\vert \Lambda \vert}}{ 4 \kappa_{1}}
\mbox{Exp} [-\frac{2 a}{\kappa_1} \sqrt{\frac{\vert \Lambda \vert}{6}}] \right ) \right ] 
=\mbox{Int} \left [ \frac{\sqrt{6}\gamma_F}{4} \nu \exp{(-2 \nu)} \right ]
\end{eqnarray}

In Table (II) we show some values of $\mu_{max}$ and $n_{max}$  with increasing
strength of Yukawa interaction ($\gamma_F$)
and varying warp-factor parameter $\nu$.

\begin{table}[htb]
\renewcommand{\tabcolsep}{1.25pc} 
\begin{tabular}{|c|c|c|c|c|c|c|c|} \hline
Yukawa & Warping & \multicolumn{2}{c|}{$\gamma_{F} = 50$} & \multicolumn{2}{c|}{$\gamma_{F} = 100$} & \multicolumn{2}{c|}{$\gamma_{F} = 150$} \\ 
\cline{3-8}
Coupling & Parameter, $\nu$ &  $\mu_{max}$  &  $n_{max}$  &  $\mu_{max}$  &  $n_{max}$  &  $\mu_{max}$  & $n_{max}$ \\ \hline
 & 0.5 & 35.81 & 8 & 71.62 & 16 & 107.43 & 24 \\ \cline{2-8}
$F(\Phi) = \Phi$ & 1 & 16.33 & 3 & 32.65 & 7 & 48.98 & 10 \\ \cline{2-8}
& 1.25 & 11.03 & 2 & 22.06 & 4 & 33.09 & 6 \\ \cline{2-8}
& 2 & 3.39  & 0 & 6.78 & 0 & 10.18 & 1\\ \hline
 & 0.5 & 30.32 & 6 & 60.65 & 11 & 90.98 & 17 \\ \cline{2-8}
$F(\Phi) = \sin \Phi$ & 1 & 18.39 & 4 & 36.79 & 8 & 55.18 & 12 \\  \cline{2-8}
 & 1.25 & 14.32 & 3 & 28.65 & 6 & 42.98 & 9 \\ \cline{2-8}
 & 2 & 6.76 & 1 & 13.53 & 2 & 20.30  & 3\\ \hline
\end{tabular}

\caption{The maximum bound state mass, $\mu_{max}$ and number of states, $n_{max}$ are 
tabulated using $\nu =0.5, 1$ and $1.25$ for increasing
values of $\gamma_{F}$ for the two different types of Yukawa coupling. As the
coupling becomes stronger the value of $\mu_{max}$ and $n_{max}$ increases, which imply that 
the possibility of localisation is more pronounced for larger $\gamma_{F}$.}
\end{table}

From the above numerical values for $\mu_{max}$ and $n_{max}$ 
we can make some comparative statements on the role of the 
two different Yukawa couplings.

(i) Equating the $\mu_{max}$ or $n_{max}$ for the two Yukawa
couplings we obtain $\nu=.8$. Thus for $\nu<.8$ the $\Phi$
coupling has a chance of having more bound states compared to
the $\sin \Phi$ one. The converse is true for $\nu>.8$. Note that
for $\nu=1$ (which is close to 0.8) the difference between the
values of $\mu_{max}$ and $n_{max}$ for the $\Phi$ and $
\sin \Phi$ couplings is small. 

(ii) In the $F[\Phi]=\Phi$ case, the factor of $\frac{\pi}{2}$ 
in the expressions for $\mu_{max}$ and $n_{max}$ and a corresponding
factor of one for the $F[\Phi]=\sin \Phi$ case seems to suggest the
following generalisation for arbitrary $F[\Phi]$ :

\begin{eqnarray}
\mu_{max} = \gamma_F \{ F[\Phi(\infty)]\} \exp {(-\{ F[\Phi(\infty)]\}\nu)}
\nonumber \\
\mbox{n}_{max} = =\mbox{Int} \left [ \frac{\sqrt{6}\gamma_F}{4} 
\left (\{ F[\Phi(\infty)]\} \exp {(-\{ F[\Phi(\infty)]\}\nu)} \right )^2 \right]\\
\end{eqnarray}

where $\Phi(\infty)$ is the asymptotic value of the kink profile,
which, in our case here is $\pi/2$.
Thus, in addition to the parameters $\gamma_F$ and $\nu$, 
one can tune the number of bound states using different choices for 
$F[\Phi]$ too.  

(iii) Also note that for $\nu =2$ or larger the $n_{max}$ and $\mu_{max}$ 
both become relatively smaller. However, as we shall see later, the
lifetime of these states (with large $\gamma_F$ and $nu\ge 2$)
is exponentially large. Therefore, they seem to be better candidates for
describing localised massive fermions.

\section{Beyond the H.O. approximation, the lifetime
of quasibound states} 

In the above section we dealt with the harmonic oscillator approximation
to get the spectrum of the lowest states. The spectrum however will be
corrected by the presence of higher order terms. In particular, as we 
shall see below, the energy eigenvalues are in general complex with the
imaginary part being related to the lifetime of the quasibound state.

To analyse this in some generality let us go back to the original
Schrodinger like equation written using the variable $\sigma$. 
One may easily recast things using $y$ or $\hat y$.
Expanding all terms in Taylor series upto order $\sigma^4$ we obtain

\begin{equation}
\partial_{\sigma}^2 \hat \xi_{L}(\sigma) + 
\left[ m^2 + g'\vert_{\sigma = 0} - \left \{ a_1 \sigma^2 - a_2 \sigma^4\right \}
\right]\hat \xi_{L}(\sigma) = 0
\end{equation} 

where,

\begin{eqnarray}
a_1 & = & \left[ m^2 f'' - g'^2 + \frac{g'''}{2}\right]_{\sigma = 0}   \\
a_2 & = & \left[ m^2 \left( \frac{f''''}{12} 
+ \frac{f''^2}{2} \right) + \frac{g'''''}{4!}
- \frac{g' g'''}{3} \right]_{\sigma = 0}
\end{eqnarray} 

In the above expressions the prime denotes derivative with respect to $\sigma$ and 
$g(\sigma) = \eta_{F}F(\Phi) + f'(\sigma)/2$. For the constant $a_1$ and $a_2$ 
positive we can treat the potential as harmonic oscillator potential
plus a perturbation term proportional to $\sigma^4$ if the parameter $a_2$ can be
assumed as very small compared to $a_1$. Given the symmetry of the system under the transformation 
$\sigma \rightarrow - \sigma$ we put a restriction on the Yukawa coupling term 
$F(\Phi)$ such that the even order derivatives of $g(\sigma)$ vanish at the location
of the brane. In addition, keeping in mind the problem in hand, 
the odd order derivatives of
f($\sigma$) is also assumed to be zero at the brane location. 
It is worth noting that under these assumptions and within the
approximation the potential has a volcano 
profile. Keeping further terms will not alter
the nature of the potential to a very great extent. Let us now find out
the lifetime of the quasi-bound state for the particular choice of 
coupling 
$F(\Phi) = \sin \Phi$. All the above mentioned  properties are satisfied 
by the specific f($\sigma$) and g($\sigma$) discussed earlier in this
article. We assume $\vert \Lambda \vert = 6$ for convenience.
In particular, we have, 

\begin{equation}
g(\sigma) =\left( \eta_{F} + \frac{1}{2} \right) \tanh \left(\frac{\sigma}{\nu} \right)
\end{equation}

Using the general formulae for $a_1$ and $a_2$ we obtain:

\begin{eqnarray}
a_1 = \left[-\frac{m^2}{\nu} + 
\frac{{\bar \eta_{F}}^2}{\nu^2} +
\frac{\bar \eta_{F}}{\nu^3} \right] \\
a_2 = \left[ \frac{m^2}{2\nu^2} \left( 1+\frac{1}{3\nu}\right ) +
\frac{2}{3} \frac{ \bar \eta_{F}}{\nu^4} \left(\frac{1}{\nu} + \bar \eta_{F} 
\right)\right] 
\end{eqnarray} 

where $\left( \eta_{F} + \frac{1}{2} \right)$ is replaced by $\bar \eta_{F}$.
It is useful to mention here that factors of $sqrt{\vert\Lambda\vert}/6$
which have been taken to be one with the choice $\vert\Lambda\vert =6$
provide the correct dimensions for each term in the above and in 
the subsequent discussions.
We note that in the range of the warping parameter $\nu \ge 2$ 
the perturbation constant $a_2$ is reasonably small compared to a. 
One can see this by further approximating the potential for 
moderately large $\nu$ and large $\bar \eta_F$. Ignoring terms 
proportional to $m^2$ and $\bar \eta_F$ we find that the
effective potential can be written as :

\begin{equation}
V(\sigma)= \frac{{\bar \eta_F}^2}{\nu^2} \left ( \sigma^2-\frac{2}{3\nu^2}\sigma^4\right )
\end{equation}

Using the above approximate potential we can use the method developed in
{\cite{zam}} for the calculation of the lifetime,
$\tau = -\frac{1}{2 ImE}$ (where ImE is a small negative quantity).
We briefly summarize their method below.
 
The imaginary part $Im E$ is given by the formula :

\begin{equation}
\mathrm{Im E = \frac{1}{i}
\frac{\displaystyle{\int_{-\infty}^{\infty} \left( \xi \frac{d^2
\xi^{\ast}}{d \sigma^2} - \xi^{\ast} \frac{d^2
\xi}{d \sigma^2} \right) d\sigma}}{\displaystyle{\int_{-\infty}^{\infty}} \vert
\xi(\sigma) \vert ^2 d\sigma}}
\end{equation} 

The denominator is calculated using 
the zeroth order wavefunction (harmonic oscillator)
in Rayleigh Schr\"{o}dinger perturbation theory. In such an approximation
we assume that the dominant contribution to the norm comes from the 
small values of $\sigma$. For the numerator, we use the WKB
method with the assumption that the probability current is dominated by
the contribution of the wave function in the classically forbidden
region. Same normalisation for the small and large $\sigma$ solution
has been guaranteed by the asymptotic matching of the
respective solutions in these regions. 

It is easy to show (following \cite{zam}) that if $\bar\eta_F$ is large and $\nu \ge 2$
the lifetime of the metastable lowest state of the harmonic oscillator 
goes as 

\begin{equation}
c \tau \approx \frac{\nu^{\frac{1}{2}}}{{\bar\eta_F}^{\frac{3}{2}}} \exp (\bar\eta_F \nu)
\end{equation}

From the above formula it is clear that the lifetime is exponentially
enhanced for large $\bar\eta_F \nu$. Note that $\nu\ge 2$
is required in order to treat the $\sigma^4$ term as a perturbation.
If $\bar\eta_F \nu \tilde > 50$ then the lifetime will be close to the
age of the universe ($10^10 years$). Therefore large $\bar\eta_F$ would mean 
values greater than 25 (for $\nu=2$). For such values of the parameters
the massive mode can be, more or less assumed to be stable.

One can also evaluate the lifetime for the $\Phi$ Yukawa coupling
though qualitatively the results will not differ except for
factors here and there. 

\section{Conclusions}
Let us now summarise pointwise the results obtained in this article
and briefly discuss some possible avenues of future research.

(i) In the presence of a sine-Gordon potential we have first discussed
an exact solution for the warp factor and the scalar kink by solving
the full set of Einstein--scalar equations in the presence of
a negative cosmological constant. The background spacetime
however, in this case, has a non--constant Ricci scalar which is
asymptotically negative. The matter stress energy which generates
this warped line element satisfies the null energy condition with 
isotropic negative pressure in the vicinity of the brane. The pressure
along the extra dimension is, however, zero at the brane location
and positive away from the brane.  

(ii) Fermion fields have been studied in the background geometry of
the exact solution mentioned in (i) above. We couple the scalar field
to the spinor field through a Yukawa coupling and write down, following
standard methods, the effective 
Schrodinger like equation for massive fermions. Two different types of Yukawa 
interactions are considered here, for each of which we analyse the issue of
localisation of massive fermions. The difference between the effective
potentials of the two Yukawa interactions lie in the shape and depth of the
well at the brane location---for the same value of the Yukawa coupling
parameter $\gamma_F$ the one which is more steep is prone to
have more bound states. Using approximate methods we are able to
obtain the maximum mass, the low-lying mass spectrum and the maximum number
of trapped states for each case, for large values of the Yukawa coupling
parameter. We also calculate the lifetime of these
quasilocalised states and show that it can indeed be exponentially large
for large $\gamma_F$ and moderate values of $\nu$. 
In addition, we also show that for zero or small values of
$\gamma_F$ (fixed $\nu$) very few bound states may exist and the spectrum
becomes almost continuous beyond these discrete states---a fact 
demonstrated through the vanishing of the level spacing for large
values of the quantum number $n$.  
		  
For massless fermions, we verify previously known results. Only the left or the
right chiral fermions can be localised under special restrictions on 
the value of the Yukawa coupling constant and other parameters that
appear in the warp factor/kink solution. 

(iii) An obvious question related to the background geometry 
arrived at in this paper is : is gravity localised on the brane
in this model? The answer to this question, as far as we have
been able to analyse, is in the affirmative. The potential felt
by the graviton fluctuations is similar to the one in the RSII model
with a well in the neighborhood of the brane followed by a barrier and
a decay to zero as we go deep into the bulk. We shall elaborate on the
localisation of gravity in our model in a future publication. 

\section*{Acknowledgments}
SK thanks S. Randjbar Daemi for discussions and the ASICTP, Trieste, Italy
for support during part of the period when this work was done. RK thanks
IIT Kharagpur for support. 
 
\maketitle

\end{document}